# Synthesis and Microstructural Studies of Iron Oxypnictide LaO$_{1-x}$F$_x$FeAs Superconductors

Chandra Shekhar[1,3], Sonal Singh[1], P K Siwach[2], H K Singh[2] and O N Srivastava[1]

[1]*Centre of Advanced Studies for Physics of Materials, Department of Physics, Banaras Hindu University, Varanasi- 221005, India*

[2]*National Physical Laboratory, Dr K S Krishnan Road, New Delhi-110012, India*

**Abstract**

We report on the synthesis and structural/ microstructural studies of iron based fluorine doped LaOFeAs superconductors. We have successfully synthesized fluorine doped superconducting LaO$_{1-x}$F$_x$FeAs materials by choosing lower temperature (∼1150$^0$C) and longer synthesis duration (∼60 hours) as compared to the standard values of these employed in the pioneering first contribution [Kamihara et al 2008 *J Am Chem Soc* 130 3296]. Decrease of lattice parameters as determined by x–ray diffraction confirm the substitution of fluorine. The superconducting transition temperature is 27.5 K which is observed at doping level of x=0.2. This superconducting material LaO$_{1-x}$F$_x$FeAs exhibits interesting microstructural characteristic. This relates to the existence of another structural phase, besides the standard phase, having *c* parameters of ∼ 12.67Å. This suggests existence of modulated structure, similar to the cuprates, in these new oxypnictides. This phase may have new impacts on this new high–T$_C$ family.

―――――――――――――――
[3]Electronic mail: cgsbond@gmail.com

**Introduction**

Since the discovery of superconductivity at 3.2K in iron-based LaOFeP compound [1], extensive efforts have been devoted towards searching new superconductors in this system. It has seen the light of the day when a team led by Hosono at the Tokyo Institute of technology [Japan], replaced P atom by As atom together with substitution of oxygen with fluorine. The resultant compound LaO$_{1-x}$F$_x$FeAs (x=0.11) shows the superconducting transition temperature (T$_C$) at 26 K [2]. Subsequently superconductivity at 25 K was also observed by partial substitution of La atom by Sr atom [3]. Shortly after this discovery, T$_C$ was surprisingly increased to more than 40 K when La in LaO$_{1-x}$F$_x$FeAs was replaced by other rare earth elements such as Ce [4], Pr [5], Nd [6], Sm [7] and Gd [8].

The compound LaOFeAs is an equiatomic quaternary of ZrCuSiAs type tetragonal layered structure with lattice parameter *a*=4.035Å, *c*=8.739Å [9] and its structure belongs to the P4/nmm space group. The crystal is composed of a stack of alternating LaO and FeAs layers. The LaO layer is sandwiched between FeAs layers. It is thought that these two layers are, positively and negatively charged respectively, and that the La–O chemical bond in the LaO layer is ionic whereas the Fe–As has a predominantly covalent nature. Thus, the chemical formula may be expressed as (La$^{+3}$O$^{-2}$)$^{+1}$ (FeAs)$^{-1}$. The charge carriers have been increased by substitution of the O$^{-2}$ ion by F$^{-1}$ ion. The parent material LaOFeAs is non-superconducting but shows spin density wave instability in between 150−160K in both resistivity and d.c. magnetic susceptibility [2, 8]. The spin density wave instability has been found to relate to structural transition from tetragonal to monoclinic [10]. Doping the system with fluorine suppresses both the magnetic order and the structural distortion in favour of superconductivity. Another important characteristic associated with this new superconductor is layered structure and hence this possesses high value of upper critical field is in these type of superconductors [11-13]. This leads to possibility of carrying high current capacity. The structural and microstructural features of this new family of superconductors have not been investigated in detail so far. Understanding of structural features is expected to assist further tailoring of this oxypnictide. We have, therefore, focused our investigation on structural and microstructural studies.

**Experimental Details**

In the present study the synthesis of F doped $LaO_{1-x}F_xFeAs$ ($0 \leq x \leq 0.4$) high temperature superconductor has been carried out by two step solid state reaction at ambient pressure. In the first step, for preparation of LaAs, $Fe_2As$ and FeAs, we mixed La (99.9% purity, 0.5–1 mm size, Leico), Fe (99.98% purity, 0.2–0.5 mm, Aldrich) and As (99.999% purity, Lump, Alfa-Aesar) in a ratio of 1:3:3 with the help of agate & pestle. The mixture powder was pelletized and then sealed in evacuated quarts tube in Ar atmosphere. The sealed silica tube was heated $900^0C$ for 12 hours. In the second step, the mixture of LaAs, $Fe_2As$ and FeAs were mixed with dehydrated $La_2O_3$ (99.99% purity, 0.1–0.2 mm size, Aldrich), La and $LaF_3$ (99.9% purity, 0.1–0.2 mm size, Aldrich) in stoichiomentry ratio. The final stoichiometry is $(1+x)La+(1-x)La_2O_3+xLaF_3+3FeAs$, $x=0$ for pure and $x=0.05, 0.1, 0.2$, for fluorine doped samples. After the final grinding, the powder was again pelletized at a pressure of 4tons/inch$^2$. The quartz tube was evacuated up to $10^{-5}$ torr and sealed. The sealed quartz tube was heated again at $1150^0C$ for 60 hours followed by furnace cooling to room temperature. We have chosen comparatively lower temperature ($1150^0C$ instead of $1250^0C$) and longer synthesis duration (60 hrs instead of 40 hrs) to avoid explosion. This is some what different than the standard synthesis temperature and duration so far adopted [2]. All the grindings have been carried out in a glove box containing $P_2O_5$, NaOH and under argon atmosphere. All the samples in the present investigation were subjected to gross structural characterization by x-ray diffraction (XRD, PANanalytical X'Pert PRO, $CuK_\alpha$ radiation), electrical transport measurements by four probe technique (Keithley Resistivity-Hall setup), surface morphological characterization by scanning electron microscope (SEM, Philips XL-20), and the microstructural characterization by High Resolution Transmission electron microscope (HRTEM, FEI, Tecnai $20G^2$). The elemental analysis has been carried out by energy dispersive analysis of x–ray (EDAX) microanalysis system which is attached with HRTEM.

**Results and Discussion**

The as synthesized samples having various doping concentration of fluorine were subjected to gross structural characterization employing x–ray diffraction technique. The XRD patterns of $LaO_{1-x}F_xFeAs$ (x=0.0, 0.1, 0.2) samples are shown in figure 1. These reveal that the synthesized material correspond to tetragonal LaOFeAs phase. The XRD analysis using a computerized program based on a least square fitting method gives lattice parameters $a$=4.039Å & $c$=8.742Å and $a$=4.030Å & $c$=8.716Å for pure LaOFeAs and doped (x=0.2) samples respectively. It is very close to the reported standard lattice parameter values [2, 9]. However, these parameters are somewhat smaller (~$a$≈0.05%, $c$≈0.27%) than the reported standard values [9]. The XRD patterns indicate that all samples have the standard LaOFeAs structure with some minor impurity phases.

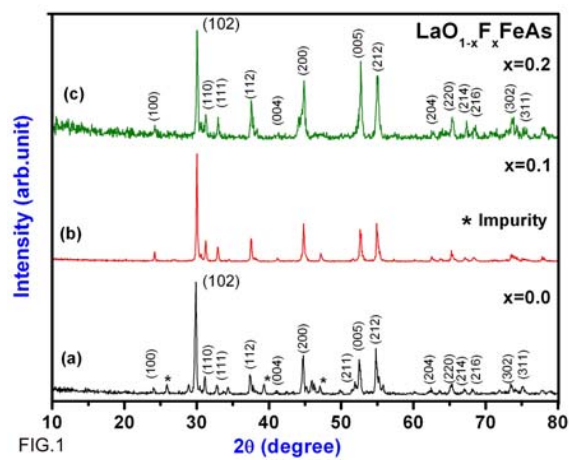

**Figure. 1.** XRD pattern of (a) LaOFeAs (b) x=0.1 fluorine doped LaOFeAs (c) x=0.2 fluorine doped LaOFeAs samples. All the indexed peaks corresponds to LaOFeAs and peaks marked by asterisk (*) are impurity phases.

Figure 2 shows the resistivity vs temperature behavior of pure and doped (x= 0.1, 0.2) samples monitored by the standard four-probe method. The resistivity of LaOFeAs shows an anomaly at 155K, which is similar to that of other reports, where, it has been shown to occur due to spin density wave instability [2, 6, 10]. The $T_C$ of the sample $LaO_{0.8}F_{0.2}FeAs$ is 27.5($\pm$0.2) K which is reproducible and slightly higher (~5.8%) in comparison to other report [2]. This may be explicable in terms of enhanced chemical pressure originating from shrinkage of lattice as brought out by comparatively smaller lattice parameters of the phase synthesized in the present case.

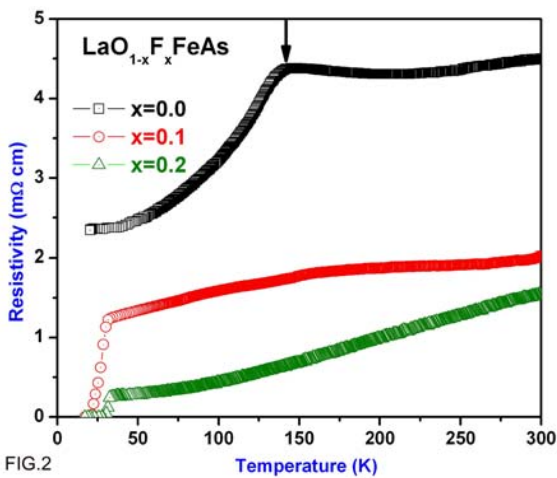

*Figure.2.* Resistivity vs temperature behaviour of pure and fluorine doped LaOFeAs samples. The superconducting transition temperature 27.5K corresponds to x=0.2 composition. The pristine sample exhibiting spin density wave anomaly marked by arrow but no superconducting transition is shown by upper curve.

The stoichiometry of various compositions of $LaO_{1-x}F_xFeAs$ was investigated by employing EDAX microanalysis system at several points. It was found that samples were homogeneous and for the specific sample nearly same stoichiometry was found at different regions. The representative example of $LaO_{0.8}F_{0.2}FeAs$ stoichiometry as determined and shown in figure 3 which approximates quite reasonably to the synthesized composition. It can thus be said that the synthesis has actually led to the envisaged composition.

Although several studies have been made in regard to the occurrence of superconductivity in fluorine doped LaOFeAs (i.e. $LaO_{1-x}F_xFeAs$), hardly any of the studies have focused on the microstructural aspect. Similarly, variations of microstructural details for optimally doped LaOFeAs have not been studied earlier. This communication is centered on the studies of the microstructural characterization of the above type of $LaO_{1-x}F_xFeAs$ superconductors. As in known the microstructural and related structural characteristics have considerable effect on the superconducting behavior. In view of this, present studies are devoted to investigations of microstructural and related structural characteristics of the new superconducting material $(LaO_{1-x}F_x)(FeAs)$.

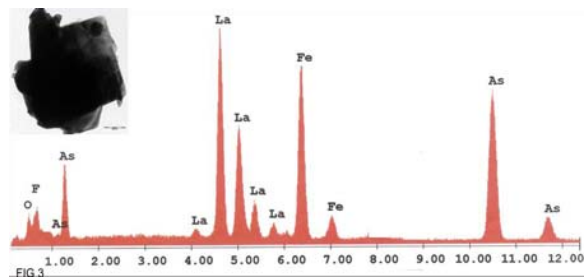

*Figure 3.* EDAX spectra of $LaO_{0.8}F_{0.2}FeAs$ composition (crystal is shown in inset).

Investigations of microstructural characteristics employing TEM (imaging and diffraction modes) explorations have revealed interesting microstructural features. The specimens for transmission electron microscope have been prepared by (a) scrapping particles from the surface of the $(LaO_{1-x}F_x)(FeAs)$ pellets (b) turning pellet into fine particles and mounting such particles which floated on benzene. These particles were mounted on holy carbon grids. The broad microstructural/ structural details for the sample prepared by (a) and (b) were found same. Thus it can be said that the microsructural features observed through TEM are representative of the superconducting phase $(LaO_{1-x}F_x)(FeAs)$. TEM exploration studies were employed for several samples of the superconducting material. Representative transmission electron micrographs of superconducting specimens are shown in figure 4.

In order to explore the structural aspects of the as grown phase, selected area diffraction patterns (SAD) particularly with electron beam along [001] direction. Representative examples of the (hk0) diffraction patterns from the $(LaO_{1-x}F_x)(FeAs)$ with x=0.2 are depicted in figure 4 (b). This diffraction pattern is in (hk0) orientation and corresponding microstructure is shown in figure 4(a). As can be seen from figure 4 (b) the diffraction spots are arranged on square grid corresponding to (100) and (010) spots. The indexing should be spacing of 4.03Å which is the expected a spacing of lattice parameter $a$ of LaOFeAs material. Further, we have taken SAD patterns with the electron beam along [100] or [010] direction. A representative diffraction patterns is shown in figure 4[d]. This figure reveals some interesting characteristics. These are (a) strong 00ℓ type diffraction spots whose indexing is outlined in figure and (b) comparatively weak spots indicative of different $c$ lattice parameter than those represented by strong diffraction spots. The analysis of bright 00ℓ spots revealed the standard $c$ spacing of ~ 8.73Å. However, the faint spots in conjunction with bright spots some of which are marked by arrows, exhibited spacing, this invariably was found to be ~ 12.67Å. A spacing of this type is shown by vertical arrows in figure 4(c). In order to get further insights relating to the occurrence of new ~ 12.67Å spacing, HRTEM micrographs were taken. A typical HRTEM micrograph is shown in figure 4(c). Careful analysis of lattice fringes has shown dominantly the presence of regular $c$ lattice parameter of ~ 8.73Å. However, the lattice fringes with ~12.67Å lattice parameter were also visible. Some such fringes revealing $c$ lattice parameter of ~ 12.67Å are marked by arrows in figure 4(c). Together with the existence of new local structure with $c$ spacing of ~ 12.67Å, staking faults were also found to be present. Some of these marked by SF in figure 4(c). It is interesting to find that the spacing ~ 12.67Å is equal to the $c$ parameter (~ 8.73Å) of the known phase of superconductor $(LaO_{1-x}F_x)$ (FeAs) plus the thickness of FeAs block (3.94Å). It can thus be taken that in addition to the known structure another structure with $c$ parameter of ~ 12.67Å representing a new phase with $c_{new}$ ~ 12.67Å also exits. The existence of this phase has been confirmed by SAD and HRTEM. The observations suggest and interesting feature of the new superconducting material $(LaO_{1-x}F_x)$ (FeAs).

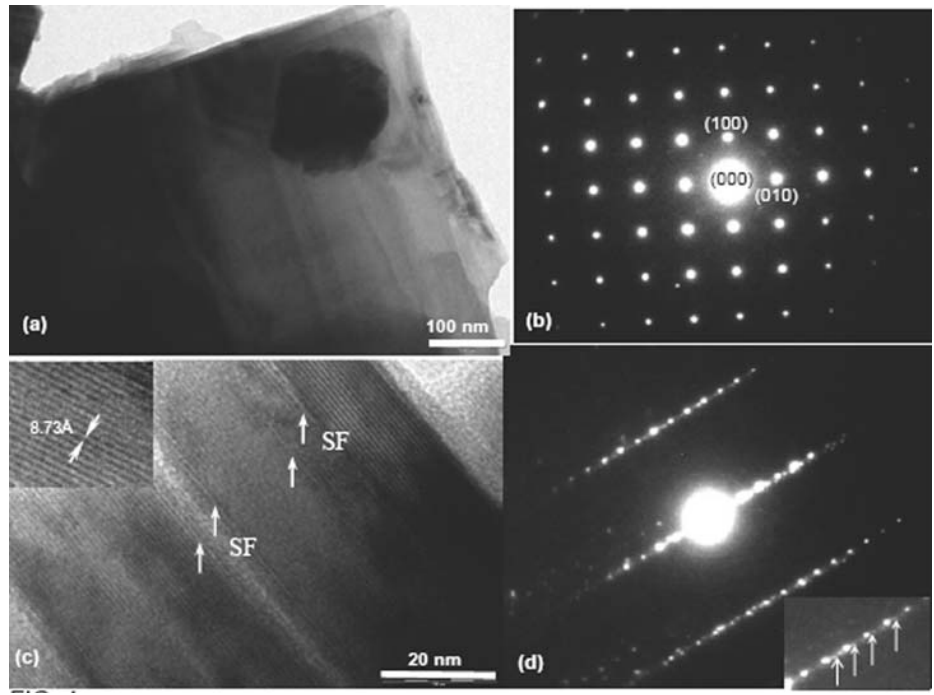

**Figure 4.** TEM micrographs of $LaO_{0.8}F_{0.2}FeAs$ sample. **(b)** SAD pattern corresponding to the microstructure (a) with electron beam along [001] direction. **(d)** SAD pattern corresponding to the microstructure (c) with electron beam along [100] or [010] direction. The lattice fringe width ~12.67Å is marked by vertical arrows in (c).

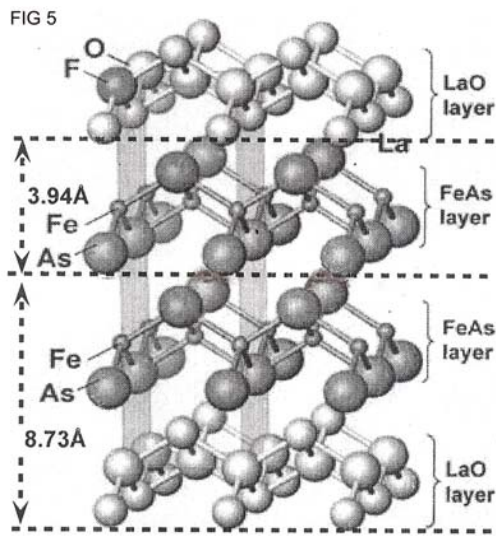

*Figure 5.* Schematic diagram of modified crystal structure of LaOFeAs. The modified lattice parameter $c_{new}$ is equal to 12.67 Å.

This relates to existence of new structural phase in $(LaO_{1-x}F_x)(FeAs)$ where $c$ parameter is equal to the standard $c$ parameter plus the thickness of the block containing charge carriers. This is similar to cuprate where Bi, Tl and Hg bearing cuprates exhibit such structural phases [14, 15]. For cuprates this block is $CuO_2$ and for the new $(LaO_{1-x}F_x)(FeAs)$, it is FeAs. A schematic figure exhibiting the new structural phase as suggested has been shown in figure 5. If the phase $(LaO_{1-x}F_x)(FeAs)$ is represented by numerical symbol 11, the new phase with two FeAs layers corresponds to 12. It should be pointed out that similar to the cuprates, for these new superconductors the transition temperature may vary for the above said different structural phases. Further investigations on this aspect are required.

**Conclusion**

Based on the present investigations it can, therefore, be concluded that for the new superconducting material $(LaO_{1-x}F_x)(FeAs)$ exhibits interesting microstructural features. This relates to the existence of another structural phase, besides the standard phase, having $c$ parameters of ~12.67Å. This is equal to the standard $c$ parameter of ~ 8.73Å and width of FeAs block (~ 3.94Å). This modified structural phase may affect the superconducting transition temperature. The existence of this new structural phase with prolonged $c$ parameter (~12.67Å) may throw new light on the superconducting characteristics of the oxypnictide family of superconductors.

**Acknowledgement**

The authors are grateful to Prof. A.R. Verma, Prof. C.N.R. Rao (FRS), Prof. D.P. Singh for their encouragement and Prof. H Kishan & Dr. V.P.S. Awana, NPL, New Delhi for fruitful discussion and suggestions. Financial supports from UGC (SUC programme) and DST-UNANST are gratefully acknowledged. One the author (Chandra Shekhar) is also gratefully acknowledged to UGC for awarding of Dr D. S. Kothari postdoctoral fellowship.